\documentclass[cite,varenna]{cimento}
\usepackage{graphicx}  
\usepackage{amsmath}
\usepackage{amsfonts}
\usepackage{amssymb}

\title{Gravitational waves, diffusion and decoherence}

\shorttitle{Gravitational waves, diffusion and decoherence}

\author{S. Reynaud and B. Lamine}

\institute{Laboratoire Kastler Brossel, ENS, UPMC, CNRS, \\ 
Université Pierre et Marie Curie, Paris 75252 FRANCE}

\author{M.-T. Jaekel}

\institute{Laboratoire de Physique Théorique, ENS, UPMC, CNRS, \\ 
Ecole Normale Supérieure, Paris 75231 FRANCE}

\shortauthor{S. Reynaud, B. Lamine and M.-T. Jaekel}

\def\beq{\begin{eqnarray}}
\def\eeq{\end{eqnarray}}
\def\beqn{\begin{equation}}
\def\eeqn{\end{equation}}

\def\e{\mathrm{e}}     
\def\r{\mathrm{r}}     
\def\s{\mathrm{s}}     
\def\d{\mathrm{d}}     
\def\u{\mathrm{u}}     
\def\m{\mathrm{m}}     
\def\K{\mathrm{K}}     
\def\Hz{\mathrm{Hz}}   

\def\bx{\mathbf{x}}     
\def\bk{\mathbf{k}}     
\def\bn{\mathbf{n}}     
\def\bp{\mathbf{p}}     
\def\bq{\mathbf{q}}     
\def\TT{\mathrm{TT}}
\def\gw{\mathrm{GW}}
\def\Sagnac{{\rm Sagnac}}

\def\at{{\rm at}}

\def\dd{\ensuremath{\mathrm{d}}}

\def\mP{m_{\mathrm{P}}}
\def\lP{\ell_{\mathrm{P}}}
\def\tP{t_{\mathrm{P}}}
\def\lC{\ell_{\mathrm{C}}}
\def\kB{k_{\mathrm{B}}}

\def\be{\mathbf{e}}     
\def\proj{\Pi}     

\begin{document}

\maketitle

\begin{abstract}
The quite different behaviors exhibited by microscopic and macroscopic systems with 
respect to quantum interferences suggest that there may exist a naturally frontier 
between quantum and classical worlds. 
The value of the Planck mass (22$\mu$g) may lead to the idea of a connection
between this borderline and intrinsic fluctuations of spacetime.
We show that it is possible to obtain quantitative answers to these questions 
by studying the diffusion and decoherence mechanisms induced on quantum
systems by gravitational waves generated at the galactic or cosmic scales.
We prove that this universal fluctuating environment
strongly affects quantum interferences on macroscopic systems, 
while leaving essentially untouched those on microscopic systems.
We obtain the relevant parameters which, besides the ratio of the system's mass 
to Planck mass, characterize the diffusion constant and decoherence time. 
We discuss the feasibility of experiments aiming at observing these effects
in the context of ongoing progress towards more and more sensitive 
matter-wave interferometry.
\end{abstract}

\section*{Introduction}


These notes correspond to the second and third lectures in a series of three 
given during the \textsc{International School of Physics ``Enrico Fermi''} on 
\textsc{Atom Optics and Space Physics} held at Varenna in July 2007.
Whereas the first lecture~\cite{Varenna1} was devoted to the discussion of the 
present status of tests of general relativity, 
the second and third ones discuss consequences of GR in terms of fluctuations 
of spacetime and of the associated diffusion and decoherence processes.

The idea that fluctuations of spacetime can be thought of as a natural source 
of decoherence that would define an ultimate border for quantum interferences
was evoked long ago by Feynman~\cite{Feynman}.
The main motivation for such an idea is the fact that the Planck mass $\mP$ built on
the Planck constant $\hbar$, the velocity of light $c$ and the Newton constant $G$, 
has a value on the borderland between microscopic and macroscopic masses
\beq
\mP \equiv \sqrt{\frac{\hbar c}G} \simeq 22\,\mu\mathrm{g} 
\eeq
This could be an accidental coincidence but might also be a hint that
ultimate fluctuations of spacetime are the cause of some universal
decoherence mechanism.

One may go one step further in this reasoning by recalling that Planck scales of 
time and length, typical for intrinsic fuzziness of spacetime, 
have extremely small values
\beq
\tP \equiv \sqrt{\frac{\hbar G}{c^5}} \simeq 5\times 10^{-44}\mathrm{s} \quad&,&\quad
\lP \equiv \sqrt{\frac{\hbar G}{c^3}} \simeq 10^{-35}\mu\mathrm{m} 
\eeq
In spite of these far from currently accessible scales, it could be possible
for the associated fluctuation behaviours to be different for
microscopic and macroscopic masses which correspond to a Compton wavelength
$\lC\equiv\hbar /mc$ respectively larger and smaller than the Planck length
$\lP\equiv\hbar /\mP c$
\beq
m < \mP \Leftrightarrow \lC > \lP \quad&,&\quad
m > \mP \Leftrightarrow \lC < \lP
\eeq

This qualitative idea~\cite{Karolyhazy66,Diosi89,Ellis90,Penrose96,Jaekel94}
corresponds to the possibility of a quantum/classical transition  
which would be a consequence of universal gravitational fluctuations.
Stated differently, quantum systems could undergo a decoherence process 
due to the fact that they occur in a fluctuating spacetime and not
in the commonly supposed Minkowski spacetime.
The resulting long-term diffusion of the phase and associated gravitational 
decoherence would here play the same role 
as Brownian motion which was able to reveal long term effects of
a large number of atomic collisions long before the atomic scale
typical of single collisions was observed~\cite{brownian}. 

In these lecture notes, this qualitative discussion is translated into more 
quantitative statements which can be obtained when specifying the 
quantum systems under study, here light or matter-wave interferometers, 
and the source of their decoherence, here the stochastic backgrounds 
of gravitational waves. 
Such backgrounds are the spacetime fluctuations deduced
from general relativity and the knowledge of our gravitational environment.
The form of their coupling to light or matter waves is also mastered,
so that the associated decoherence mechanism can be fully calculated and
characterized~\cite{Reynaud01,Lamine02}. 
In particular, relevant figures are extracted which depend not only on the 
comparison of the mass with the Planck mass, but also on the geometry of 
the quantum interferometer and on the gravitational noise level
determined by the GW background~\cite{Reynaud04,Lamine06}.

This discussion has to be considered in the more general study of
quantum decoherence as a universal phenomenon affecting all physical systems 
coupled to a fluctuating environment%
~\cite{Zeh70,Dekker77,Zurek81,Caldeira83,Joos85}.
This effect determines a transition between quantum and classical behaviors, 
and thus justifies a classical description of large enough systems.
Decoherence thus plays a key role in adressing the ``Schr\"odinger's 
cat problem''~\cite{Schrodinger,Zeh,DecoherencePoincare}
even if it does not solve the whole ``Quantum Measurement'' problem%
\cite{WheelerZurek}. 
For any kind of environment, not necessarily of gravitational origin, 
quantum decoherence is very rapid for large macroscopic systems, 
while remaining inefficient for small microscopic ones.
This phenomenon has been observed on ``mesoscopic'' systems for 
which the decoherence time is neither too long nor too short, such as
microwave photons stored in a high-Q cavity \cite{Brune96} 
or trapped ions \cite{Myatt00}.
In such model systems, the environmental fluctuations are particularly
well mastered and the quantum/classical transition has been shown 
to fit the predictions of decoherence theory \cite{Raimond01}.

In the following, we will describe our gravitational environment,
which consists of stochastic gravitational waves generated
at galactic or cosmic scales.
To this aim, we will first present an introduction to gravitational
waves and their effects on the observables met in light or
matter interferometers.
This vast domain has been the topic of a number of books 
(see for example~\cite{Landau,Weinberg,Misner,Blair,Tourrenc,EhlersVarenna}) or reviews 
(see for example~\cite{Thorne,Hellings,Schutz,Maggiore,Blanchet,Armstrong}).
The present lecture notes will be focused on the discussion of topics 
selected for their direct connection with the questions treated at this School, 
in particular the effect of gravitation waves on optical or 
matter-wave interferometers~\cite{GWVarenna,AtomsVarenna}.

The calculations will be applied to matter-wave interferometers such as
the project \textsc{Hyper} which was designed for space experiments~\cite{Hyper00},
thus taking advantage of the quieter environment available in space.
Such instruments have their performance presently limited by instrumental dephasings 
produced for example by vibrations of the mechanical structure, residual collisions or
thermal radiation effect. In principle, these noise sources can be reduced by using 
better vacuum, lower temperature, improved velocity selection $\ldots$ 
With the ongoing progress in matter-wave interferometry, more fundamental limits may 
eventually be reached such as the gravitational decoherence. 

Larger and larger molecules are being used in matter-wave 
interferometers~\cite{Zeilinger}, and it is worth wondering whether there is a limit 
to the mass of the molecular probes beyond which interferences would no longer 
be visible.
This question has been discussed in the context of studies devoted to the ``spacetime 
foam'' resulting from some quantum gravity models~\cite{Percival,Garay98,Amelino,Bingham}.
Here it is treated for gravitational decoherence induced by the scattering 
of GW stochastic backgrounds~\cite{Reynaud01,Lamine02,Reynaud04,Lamine06}. 
As will be shown below, the mass of the probe is a relevant parameter,
but not the only one, for describing the approach to the quantum/classical border.
The nature of the coupling to fluctuations and the noise level 
also play a role in the quantitative answers to this question.

\section{Tracking observables}

When he introduced the relativistic conception of space-time~\cite{Einstein05}, 
Einstein emphasised that remote clocks have to be compared through the 
transfer of time references encoded on electromagnetic pulses.
This clock synchronisation procedure between distant observers
is nowadays a routine for metrological applications~\cite{TFmetrology} 
as well as the basic building block of Global Navigation Satellite
Systems~\cite{GNSS} or the definition of reference systems~\cite{refsystems}.
The tracking observables used in most gravity tests performed in the 
solar system are also built up on similar protocols. 
This statement is valid 
for example for radio tracking of planetary probes~\cite{radioscience}, 
in particular Pioneer~\cite{pioneer} and Cassini probes~\cite{cassini}, 
as well as lunar laser ranging~\cite{LLR}.
In the present section, we present the basic tracking observables which
are built up on electromagnetic time transfer signals exchanged between
remote observers and compared to locally available atomic clocks. 
We will see in the next section how these observables are affected by
gravitational waves. 

To make the discussion more concrete, let us consider a synchronisation
between an atomic clock on board a probe in space and another one colocated 
with a station on ground. Both clocks (observers) are supposed to be
equipped with transfer capabilities corresponding to the emission and 
reception of a time reference encoded as the ``center of energy'' of an
electromagnetic pulse or a marked phase front.
The time reference transferred from one observer to the other is by
necessity a quantity encoded on the field and preserved by the law
of propagation. 
In special relativity, one may
define this time reference as the light cone variable which 
labels the various phase fronts along the line of sight.
To be more precise, let us consider an emission event with 
positions $t_\e$ in time and $x_\e$ on the space axis along which
the transfer takes place. Then the transfer connects the reception event 
with positions $t_\r$ in time and $x_\r$ on the space axis 
with the emission event with positions $t_\e$ and $x_\e$ 
if they share the same value of the light-cone variable 
\beq
\label{Minkowski_synchro}
t_\e-\frac {x_\e}c = u = t_\r-\frac {x_\r}c 
\eeq 
This equation also means that the spatial distance $\left|x_\r-x_\e\right|$
between the two events is directly determined by the elapsed time
$c\left(t_\r-t_\e\right)$.
As a consequence, one of the observers may measure its distance to the other
one as follows~: he emits a pulse to be reflected by the remote observer
and measures the time elapsed till he gets the pulse back; he then deduces 
the spatial distance as half this lapse time; this well known principle
of radar ranging produces what is called a ``range''.

In presence of gravity fields, we have to replace the simple relation 
(\ref{Minkowski_synchro}) by a more general theoretical description
of the light cones which connect emission and reception events.
To this aim, we define the infinitesimal metric element $\d s$
from the metric tensor $g_{\mu\nu}$ characterizing space-time 
and displacements $\d x^\mu$ in space-time 
\beq
\label{metric_element}
\d s^2\equiv g_{\mu\nu}\d x^\mu \d x^\nu 
\eeq
The light cones are the solutions of the eikonal equation written
for a massless probe $\d s^2=0$. 
In the weak gravitational field of the solar system, there is a unique 
solution for an electromagnetic ray going from one point to the other.
For any couple of massive observers, labelled 1 and 2, the light-cone
solution can be written as a relation between the times 
of the emission and reception events.
This relation can be obtained in a variety of representations, 
for example through an explicit solution of the light-cone equations~\cite{explicit},
or using the so-called world function~\cite{worldfunction} or 
time delay function~\cite{JRcqg06}. In the latter point of view,
the light-cone equation is written as 
\beq
t_2-t_1=\mathcal{T} \left(t_1,\mathbf{x}_1(t_1);t_2,\mathbf{x}_2(t_2)\right)
\eeq
The timedelay function $\mathcal{T}$ measures the time taken to propagate
the pulse between the emitter and receiver, the motions of which are 
described by $\mathbf{x}_1(t_1)$ and $\mathbf{x}_2(t_2)$.

In order to illustrate the method, we write the timedelay function $\mathcal{T}$ 
in the solar system with a few simplifying assumptions~:
we ignore other gravity sources than the Sun, taken as pointlike and motionless; 
we use the Eddington gauge convention where spatial coordinates are isotropic 
(see~\cite{Varenna1} for a more detailed discussion) and treat the effect
of gravity up to first order in $GM$ where $G$ is the gravitational constant
and $M$ the mass of the Sun. 
With these assumptions (and taking $\gamma\equiv1$ for general relativity;
see discussions in \cite{Varenna1}), the timedelay is 
the following function of the spatial positions of the two endpoints
($R_{12} \equiv|\mathbf{x}_2-\mathbf{x}_1|,\,
r_1\equiv|\mathbf{x}_1|,\, r_2\equiv|\mathbf{x}_2|$)
\beq
\label{Shapiro}
&&c\mathcal{T} \left(\mathbf{x}_1;\mathbf{x}_2\right) = R_{12} + 
\frac{2GM}{c^2} \ln\frac{r_1+r_2+R_{12}}{r_1+r_2-R_{12}} 
\eeq
In this function, the first term $R_{12}$ represents the Minkowskian 
approximation (zeroth-order in $GM$), while the second term is the
(first-order) effect of gravity on electromagnetic propagation, 
known as the Shapiro time delay~\cite{Shapiro}.
The order of magnitude of this term is fixed by $2GM/c^2\simeq3$km.

At this point, it is worth emphasizing that the light-cone relation 
between emission and reception endpoints is independent of the
coordinate system (\textit{i.e.} gauge invariant) although the explicit 
expression of the timedelay function depends on the specific coordinate 
system (specific gauge convention).
There are some conveniency reasons for choosing the Eddington gauge convention
for which the spatial part of the metric is isotropic, so that the 
propagation of electromagnetic field also respects isotropy 
(with the simplifications evoked in the preceding paragraph). 
But other gauge choices would be perfectly as respectable as this one,
and they would lead to the same final expressions for the physical observables. 
This is true in particular for the range defined in the foregoing paragraphs
as a gauge invariant spatial distance between remote observers.

To fix ideas, let us suppose we study time transfer between two atomic clocks,
one brought on board a space probe, the other one colocated with a 
radio or laser station on Earth~\cite{SAGAS}. 
Such transfers can be performed on up- as well as down-links. 
The uplink signal emitted connects two points at positions 
$(t_1^\u,\bx_1(t_1^\u))$ on ground and $(t_2^\u,\bx_2(t_2^\u))$ in space
while the downlink signal connects 
$(t_1^\d,\bx_1(t_1^\d))$ on ground and $(t_2^\d,\bx_2(t_2^\d))$ in space.
When the two links are crossing at the space endpoint ($t_2^\u\equiv t_2^\d$), 
the observer on ground may define a spatial distance to the remote probe 
from the times of emission and reception at his station.
This distance is a gauge invariant observable (independent of
the coordinate system) as soon as it is written in terms of clock indications
\beq
\label{range}
\ell_{12} \equiv {\Delta s \over2} \quad,\quad
\Delta s \equiv s_1^\mathrm{d} - s_1^\mathrm{u} 
= \int_{t_1^\u} ^{t_1^\d} \d s_1 
\quad&\leftrightarrow&\quad 
\left[ \mathrm{ranging\;with\;} s_2^\d = s_2^\u\right]
\eeq 
$s_1$ is the proper time (multiplied by $c$) as measured by the clock on ground, 
and $\ell_{12}$ is half the proper time elapsed during the roundtrip
of the signal to and from the probe. 

Note that it can also be written as the half sum of quantities defined
on the up- and down-links (with $s_2^\d = s_2^\u \equiv s_2$)
\beq
\label{range_updown}
\ell_{12} \equiv {\ell_{12}^\u + \ell_{12}^\d \over2} 
\quad&,&\quad \ell_{12}^\u \equiv s_2 - s_1^\u 
\quad,\quad \ell_{12}^\d \equiv s_1^\d - s_2 
\eeq 
The latter are proper relativistic observables, which do not depend
on the coordinate system, but however depend on the choice of the origins
of proper times for the two clocks.
This dependence can be fixed by a convention stating that
the two up- and down- one-way ranges are equal, at some point chosen 
for any conveniency reason.

The Doppler tracking observables which have been over years the main source 
of information on the navigation of remote probes \cite{radioscience}
are directly related to the time derivatives of these ranges.
In order to make this point explicit, let us differentiate (\ref{range_updown})
with respect to a commonly defined time, for example $\d s_2$.
We thus get the ratios of frequencies measured on the up- and down-link
at the ground station as
\beq
\label{Doppler}
&&{\omega_1^\d\over\omega_1^\u} = {\d s_1^\u\over\d s_1^\d} =
{ 1 - \dot\ell_{12}^\u \over 1 + \dot\ell_{12}^\d  }
\quad,\quad \dot\ell_{12}^\u \equiv {\d\ell_{12}^\u\over \d s_2} 
\quad,\quad \dot\ell_{12}^\d \equiv {\d\ell_{12}^\d\over \d s_2} 
\eeq 
This Doppler observable is directly given by time derivatives of the ranges.
In the non relativistic limit, where velocities and gravity fields are
small, it can be written in terms of the range (\ref{range}) only
\beq
\label{Doppler_approx}
&&{\omega_1^\d\over\omega_1^\u} = {\d s_1^\u\over\d s_1^\d} \simeq
1 - \dot\ell_{12}^\u - \dot\ell_{12}^\d  = 1 - 2 \dot\ell_{12}    
\quad,\quad \dot\ell_{12} \equiv {\d\ell_{12} \over \d s_2} 
\eeq 
It follows that the observable $\dot\ell_{12} $
can be regarded as a properly defined relative velocity between the
two remote observers. 
Note that the Doppler observable (\ref{Doppler}) contains 
not only what is usually called the Doppler effect (relative motion 
of one observer with respect to the other) but also the Einstein effect
(effect of gravity on the clock rates) as well as the Shapiro effect
(effect of gravity on the propagation of light). 
In the non relativistic limit, it can be written as a sum of terms each
corresponding to one of these interpretations.
It must however be kept in mind that the individual terms are no
longer gauge invariant whereas the sum is gauge invariant from
its very definition.

Before ending this introduction on the status of tracking observables,
let us recall that the Doppler observable is the only source 
of information on the navigation of Pioneer probes~\cite{pioneer}
(see the discussions and references in \cite{Varenna1}) as well as 
the primary signal in the Cassini relativity experiment~\cite{cassini}.

\section{Gravitational waves (GW) in linearized general relativity}

Electromagnetic signals feel the gravitational waves (GW) along their
propagation, so that the tracking observables are affected by the 
stochastic GW backgrounds which permeate our environment and 
will be the main topic of the foregoing sections. 
We refer the reader interested in a complete description
to books such as~\cite{Landau,Weinberg,Misner,Blair,Tourrenc} 
or reviews~\cite{Thorne,Hellings,Schutz,Maggiore,Blanchet,Armstrong}.
Here, we briefly recall the main properties of GW as the free radiative solutions
of linearized general relativity, noting that the 
latter is an excellent first approximation for studying propagation
and detection of GW~\cite{GWVarenna}.

We begin by a brief reminder of this linearized theory.
The metric field is written as the sum of the Minkowski metric $\eta_{\mu\nu}$ 
(spacetime without gravity fields) and of a perturbation $h_{\mu\nu}$ 
\beq
\label{metric_perturbation}
&&g_{\mu\nu} = \eta_{\mu\nu} + h_{\mu\nu}, \qquad \eta_{\mu\nu} = {\rm{diag}}(1,-1,-1,-1)
\eeq
The law of geodesic motion is written 
\beq
\label{geodesic}
{\d u^\lambda \over\d s} = \Gamma^\lambda_{\mu\nu} u^\mu u^\nu 
\quad,\quad u^\mu \equiv {\d x^\mu \over\d s} 
\eeq
where $u$ is the proper velocity 
while the Christoffel symbols $\Gamma^\lambda_{\mu\nu} $
are defined from partial derivatives of the metric
(linearized expression keeping only first order in $h$) 
\beq
\Gamma^\lambda_{\mu\nu} = \eta^{\lambda\rho} \Gamma_{\rho,\mu\nu} 
\quad,\quad
\Gamma_{\rho,\mu\nu} &=&\frac{
\partial_\mu h_{\rho \nu} 
+\partial_\nu h_{\rho \mu} 
-\partial_\rho h_{\mu \nu}
}2 
\label{Christoffel}
\eeq
The (linearized) Riemann curvatures are obtained through a further differentiation
\beq
&&R_{\mu \rho \nu \sigma }
= \partial_\rho \Gamma_{\sigma,\mu\nu} - \partial_\mu \Gamma_{\sigma,\rho\nu} 
=\frac{
\partial_\rho \partial_\nu h_{\mu \sigma}
+\partial_\mu \partial_\sigma h_{\rho \nu} 
-\partial_\mu \partial_\nu h_{\rho \sigma}
-\partial_\rho \partial_\sigma h_{\mu \nu}
}2
\label{Riemann}
\eeq
The (linearized) Ricci and Einstein curvature tensors are deduced through contractions 
\beq
&&R_{\mu \nu} =\eta ^{\rho \sigma }R_{\mu \rho \nu \sigma }  \quad,\quad
R = \eta ^{\mu \nu }R_{\mu \nu } \quad,\quad
G_{\mu \nu} = R_{\mu \nu} - \eta_{\mu \nu} \frac R2
  \label{Ricci}
\eeq
The Einstein curvature tensor is particularized as having
a null divergence ($\eta ^{\rho \nu }\partial_\rho G_{\mu \nu } = 0 $).
The sign conventions for these definitions are those of~\cite{Landau}.

It is worth specifying the gauge transformations of these quantities, 
that is to say the change of their explicit forms under an 
infinitesimal coordinate tranformation 
\begin{eqnarray}
\label{Gauge}
x^\mu\;\rightarrow\overline{x}^\mu \equiv x^\mu - \xi^\mu(x)  
\end{eqnarray}
The metric transformation expresses the invariance of the metric element
(\ref{metric_element})
\begin{eqnarray}
&&\overline{g}_{\mu\nu} \d\overline{x}^\mu \d\overline{x}^\nu \equiv 
{g}_{\mu\nu} \d{x}^\mu \d{x}^\nu 
\;\rightarrow\;
\overline{h}_{\mu\nu} = h_{\mu\nu} + \partial_\mu\xi_\nu + \partial_\nu\xi_\mu
\quad,\quad \xi_\mu \equiv \eta_{\mu\lambda} \xi^\lambda
\end{eqnarray}
The connection transformation corresponds to the invariance of the description
of geodesic motion (or parallel transport properties)
\begin{eqnarray}
&&\overline{\Gamma}_{\rho,\mu\nu} =
{\Gamma}_{\rho,\mu\nu} + \partial_\mu \partial_\nu\xi_\rho
\end{eqnarray}
Then the Riemann curvature is gauge invariant, \textit{i.e.} has its form
preserved under the gauge transformation (\ref{Gauge}),
\begin{eqnarray}
&&\overline{R}_{\mu \rho \nu \sigma } = R_{\mu \rho \nu \sigma }
\end{eqnarray}
Using their expressions written above, it follows that the other curvature
tensors are also gauge invariant.
As repeatedly stressed in the preceding paragraphs, these relations 
are first order approximations
of more general relations of the complete geometrical theory.
They are used here for the sake of giving a simple description of GW.

Gravitational waves (GW) are the radiative solutions of Einstein equation, 
freely propagating far from their sources.
The Einstein equations are 
\beq
G_{\mu \nu }= \frac{8\pi G}{c^4} T_{\mu \nu }
\label{Gmunu}
\eeq
and GW are thus characterized in general relativity by $G_{\mu \nu }=0$ or, 
equivalently, $R_{\mu \nu }=0$. In order to write this condition,
we use (\ref{Riemann},\ref{Ricci}) to get 
\beq
&&R_{\mu \nu }={\partial_\nu \partial^\sigma h_{\mu \sigma}  
+ \partial_\mu \partial^\rho h_{\rho \nu} 
- \partial_\mu \partial_\nu h
- \square h_{\mu \nu} \over2}
\;,\quad
h\equiv \eta ^{\rho \sigma } h_{\rho \sigma}
\;,\quad
\square \equiv \eta ^{\rho \sigma } \partial_\rho \partial_\sigma 
\eeq
We can then make the particular gauge choice 
\beq
&&\partial^\rho h_{\rho \nu }=\frac{\partial_{\nu }h}2
\quad\rightarrow\quad
R_{\mu \nu }= - {\square h_{\mu \nu} \over2}
\eeq
which proves that GW are propagating 
at the speed of light, as free radiative electromagnetic waves.

In linearized general relativity as in electrodynamics, these 
solutions are conveniently described through a decomposition over plane waves
corresponding to wavevectors $k$ lying on the light cone ($k^2= 0$, that is
also $\omega \equiv ck_0 \equiv c\left|\bk\right|$)
\beq
&&h_{\mu\nu}\left( t,\bx\right) = \int \frac{\d^3 \bk}{\left( 2\pi \right) ^3}
\ h_{\mu\nu}\left[ \bk\right] e ^{ -i\omega t +i\bk.\bx } + c.c. 
\label{GWplane}
\eeq
A particularly simple description is obtained in the so-called transverse
traceless gauge (``TT'' gauge) where metric components with a temporal index 
vanish, while spatial components are transverse (with respect to the
wavevector) and have a vanishing trace 
\beq
&&h_{0\nu}^\TT \left[ \bk\right] = 0 \quad,\quad
\eta^\mathrm{ij} \bk_\mathrm{i} h_\mathrm{jl}^\TT  \left[ \bk\right] = 0 \quad,\quad
\eta^\mathrm{ij} h_\mathrm{ij}^\TT  \left[ \bk\right] = 0 
\label{TT}
\eeq
Here, bold characters represent spatial components
and latin letters spatial indices.
Note that the TT solution cannot be eliminated
through a further gauge transformation,
as proven by its direct relation to the 
Riemann curvature (\ref{Riemann})
\beq
&&R_\mathrm{i0j0} (x) = -\frac 12 \frac{\partial ^2 h_\mathrm{ij}^\TT  (x)}{c^2\partial t^2}  
\quad, \quad
R_\mathrm{i0j0} \left[ \bk\right] = \frac{k_0^2}{2} h_\mathrm{ij}^\TT \left[ \bk\right] 
\label{TTRiemann}
\eeq

The TT solution may be written as a sum over two polarizations~\cite{Blanchet}
\begin{eqnarray}
\label{TTpolar}
h_{ij}^\TT(t,\bx) &=& \int\frac{\dd^3\bk}{(2\pi)^3}
\left(\be^+_{ij}[\bk] h_+[\bk]+ \be^\times_{ij}[\bk] h_\times[\bk]
\right)
e^{-i\omega t+i\mathbf{k}\cdot\mathbf{x}}+c.c. \\
\be^{+}_{ij}[\bk] &=& \bp_i\bp_j-\bq_i\bq_j \quad,\quad
\be^{\times}_{ij}[\bk] =\bp_i\bq_j+\bq_i\bp_j \nonumber
\end{eqnarray}
The GW tensor polarizations are defined as quadratic forms of electromagnetic
polarizations, the latter being built up on two unit spatial vectors $\bp$ 
and $\bq$ orthogonal to the propagation unit vector $\bn\equiv\bk/|\bk|$ 
\beq
&&\bp_\mathrm{i}\bp_\mathrm{j}+\bq_\mathrm{i} \bq_\mathrm{j} 
= \delta _\mathrm{ij} - \bn_\mathrm{i} \bn_\mathrm{j} 
\equiv \proj_\mathrm{ij} \quad, \quad 
\bp.\bn = \bp.\bq = \bn.\bq = 0
\eeq
The dot is a scalar product for spatial vectors;
$\delta _\mathrm{ij}$ is a Kronecker symbol for spatial indices with 
$\eta _\mathrm{ij}=-\delta _\mathrm{ij}$
and $\proj _\mathrm{ij}$ a transverse projector with respect to $\bn$.
The tensor polarizations obey sum rules which will be used below to
write the noise energy
\beq
\label{projtrans}
\be _{ij}^+\be _{lm}^++\be _{ij}^\times\be _{lm}^\times =
\proj _{ijlm} \equiv
\proj _{il}\proj _{jm}+\proj_{im}\proj _{jl}-\proj _{ij}\proj _{lm}
\eeq
The metric may alternatively be decomposed over two circular polarizations, 
which leads to completely equivalent calculations.

\section{The effect of GW on tracking observables}

We now evaluate the effect of gravitational waves on the tracking observables.  
We discuss GW backgrounds as sources of noise
on the corresponding links and then present rapidly the constraints
which have been obtained from experiments on planetary probes,
particularly Cassini~\cite{Armstrong}. 

We first remind that the range as defined in (\ref{range})
directly registers the Riemann curvature.
This is the working principle of GW detectors~\cite{GWVarenna},
which can be analyzed in the simplest manner with 
GW described in the TT gauge.
We first consider two neighbouring observers with geodesic motions
and non relativistic velocities (spatial velocities much smaller
than the speed of light).
We assume linearized general relativity to be sufficient, so that
we may use a ``superposition principle'' and simply add the effects
of the GW to that of the other gravity sources (for example, the
gravity field of Earth or Sun).
With these assumptions, geodesic motions of the two observers are
unaffected by GW, no more than their clock rates
(see eqs.~\ref{geodesic}, \ref{Christoffel}, \ref{TT}).
It follows that GW only affect the value of the range (\ref{range})
through a modification of the propagation time delay of the 
electromagnetic field from one observer to the other.
The modified range has the following form, for propagation along
the line of sight here taken as the axis 1,
\beq
\ell = \left( 1-{ h_{11}^\TT\over2}\right) \ell^{(0)}
\eeq
$\ell^{(0)}$ represents the zeroth-order value of the range
(\textit{i.e.} what it would be in the absence of GW) and 
$\ell$ its perturbed value at first order in $h$.
This simple interpretation, which corresponds to 
an apparent change of the speed of light along
the line of sight, is valid only in the specific TT gauge.
It can be given an explicitly gauge invariant form by
writing the second order time derivative of the range 
(using eq.~\ref{TTRiemann} and assuming the zeroth-order range
$\ell^{(0)}$ to be time independent)
\beq
&&{\partial ^2 \ell \over c^2\partial t^2}  = -\frac12
{\partial ^2 h_{11}^\TT \over c^2\partial t^2} \,\ell^{(0)} =
R_\mathrm{0101} \,\ell^{(0)} 
\label{second_deriv_range}
\eeq
This is the law of geodesic deviation~\cite{EhlersVarenna}
written here between two neighbouring observers which would
have a constant range in the absence of GW.
It can be written in any gauge with the Riemann curvature
given by its general (linearized) expression (\ref{Riemann}).
As the expression is written at first order, the choice
of the conventional time $t$ does not matter provided
that it fits at zeroth order the clock indications of
the measuring observer. 

We now come to the study of tracking observables corresponding
to a finite (non infinitesimal) propagation time.
In this case, the range between remote observers simply 
registers the TT metric integrated along the electromagnetic links.
Using the notations introduced above, we may write it precisely
as the sum over the up- and downlinks (at first order in $h$
and assuming the zeroth-order ranges to be time independent)
\beq
&&\ell \equiv {\ell^\u + \ell^\d \over2} 
\quad,\quad 
\ell^{\u,\d} = \ell^{(0)} -
\frac12 \int_{[\u,\d]} h_{\u,\d}^\TT \d\sigma \quad,\quad 
h_{\u,\d}^\TT\equiv h_{ij}^\TT \frac{\d x_{\u,\d}^\mathrm{i}}{\d\sigma}
\frac{\d x_{\u,\d}^\mathrm{j}}{\d\sigma}
\label{finiterange}
\eeq 
The integrals run along the up- or downlinks paths [u] or [d] and
the GW amplitudes $h_\u^\TT ,h_\d^\TT $ are the projections along these
paths of the TT metric perturbations. 
In these equations, $\sigma $ is the affine parameter along the path
measured as the unperturbed range $\ell_{12}^{(0)}$, 
and $\frac{\d x}{\d\sigma }$ the
electromagnetic wavevector reduced so that its time component is unity.
As previously, the last expressions can be given an explicitly gauge 
invariant form (same assumptions as for eq.~\ref{second_deriv_range})
\beq
&&{\partial ^2 \ell \over c^2\partial t^2} =
-\frac12 \int_\u 
{\partial ^2 h_{11}^\TT \over c^2\partial t^2}  \d\sigma 
-\frac12 \int_\d 
{\partial ^2 h_{11}^\TT \over c^2\partial t^2}  \d\sigma =
\int_\u R_\mathrm{0101} \d\sigma +
\int_\d R_\mathrm{0101} \d\sigma 
\eeq 

In order to go further in the explicit calculation of these effects,
we will use the plane wave decomposition (\ref{GWplane})
of the GW perturbations.
For the sake of simplicity, we focus our attention from now on to the
case of a stationary, isotropic and unpolarized GW background 
at the classical limit with a number of gravitons per mode much
larger than unity (more discussions and references in the next section). 
We also suppose that the up- and downlinks correspond to the same
line of sight (axis 1) with opposite directions of propagation.
Speaking in terms of a Fourier decomposition in the frequency
domain, the amplitude $\delta \ell [ \omega ] $ is thus an integral over
contributions of GW corresponding to the frequency $\omega$ and 
wavevectors $\bk= |\bk|\bn $ with different
propagation directions $\bn$
\beq
&&\delta \ell[\omega]=-\frac{\omega}{2\pi c^2}
\int\frac{\dd^2\bn}{4\pi}\,h^\TT_{11}[|\bk|\bn ]\,\lambda[|\bk|\bn ]
\label{fourierq}
\\
&&\lambda[|\bk|\bn ] = \frac{ 1-e^{i\left( 1-\mu \right) \omega \tau }} {2i\left( 1-\mu \right) }  
- \frac{ 1-e^{-i\left( 1+\mu \right) \omega \tau }}{2i\left( 1+\mu \right) }  \nonumber
\eeq
The response amplitude $\lambda$ depends on $\bk$ through the frequency 
$\omega=c|\bk|$ and the cosine of the angle between the propagation directions 
of GW perturbation and electromagnetic link ($\mu\equiv \bn_1$ for a 
propagation along the axis 1). It also depends on the time of flight $\tau$ 
of electromagnetic field from one endpoint to the other.

In order to characterize the range fluctuations, we introduce the
noise spectrum $C_{\ell\ell}[\omega]$, \textit{i.e.} the Fourier transform 
of the autocorrelation $C_{\ell\ell}(t)$ of the classical stochastic 
variable $\delta\ell$
(for a classical noise, $C_{\ell\ell}(t)$ and $C_{\ell\ell}[\omega]$ are 
real and even functions)
\begin{equation}
C_{\ell\ell}(t)\equiv\left\langle \delta\ell(t) \delta\ell(0) \right\rangle 
\equiv\int_{-\infty}^{+\infty} \frac{\dd\omega }{2\pi} C_{\ell\ell}[\omega] 
e^{-i\omega t}  
\end{equation}
The evaluation of range fluctuations evoked in the preceding paragraph
thus leads to the following expression 
\beq
&&C_{\ell\ell} [ \omega ] =\frac{15c^2}{32\omega ^2} 
b[ \omega ] C_{h_{11}h_{11}} [ \omega ]  \quad,\quad
b[ \omega ] = \int \frac{\dd^2\bn }{4\pi} \left( 1-\mu^2\right)^2
\vert\lambda[|\bk|\bn ]\vert^2 
\label{spectrum}
\eeq
$C_{h_{11}h_{11}} [ \omega ]$ represents the autocorrelation 
of the metric component $h_{11}$ evaluated as a function of time 
at a fixed space position $\bx$ (more discussions in the next section)
\begin{equation}
C_{h_{11}h_{11}} (t)\equiv\left\langle h_{11} (t,\bx) h_{11} (0,\bx) \right\rangle 
\equiv\int_{-\infty}^{+\infty} \frac{\dd\omega }{2\pi} C_{h_{11}h_{11}} [\omega] 
e^{-i\omega t}  
\end{equation}
As expected from the discussion of the preceding paragraph, 
the dimensionless sensibility function $b$ is obtained by averaging 
over the direction $\bn$ the squared amplitude $\left|\lambda[|\bk|\bn ]\right|^2$ 
weighted by the noise of $h^\TT_{11}[|\bk|\bn ]$, that is $\left( 1-\mu^2\right)^2$
as a consequence of (\ref{projtrans}). 
The result of the integration over angular variables leads to~\cite{Jaekel94} 
\beq
b[\omega]= \frac{3-\cos \left(2\omega \tau \right) }{3}
- \frac{3+\cos \left(2\omega \tau \right) }{\left( \omega \tau \right) ^2}
+ \frac{ 2\sin \left(2\omega \tau \right) }{ \left( \omega \tau \right) ^3 }
\label{valb}
\eeq
When the range is much smaller than the GW wavelength ($\omega\tau\ll1$), 
a simpler expression is recovered, which corresponds
to the limit of ranging between neighbouring geodesics
\beq
b[\omega]\simeq \frac{8}{15} \left( \omega \tau \right) ^2
\quad,\quad C_{\ell\ell} [ \omega ] =\frac{c^2\tau^2}{4} 
 C_{h_{11}h_{11}} [ \omega ] \quad,\quad \omega\tau\ll1
\eeq

We end this discussion of the effect of GW on tracking observables
by recalling rapidly the constraints which have been obtained from 
tracking of various planetary probes, Voyager, Pioneer 10/11,
Ulysses, Galileo and martian probes, with the best results to date
obtained with Cassini~\cite{CassiniGW}.
The observations are reviewed in~\cite{Armstrong}, with an emphasis
on some technical issues of relevance and the associated noise analysis. 
The constraints from Cassini 2001-2002 observations,
the best obtained to date from tracking observations,
are summarized on Fig.23 of~\cite{Armstrong}.
They are given in terms of a dimensionless parameter $\Omega_\gw$
which measures GW energy density (definition below).

\section{Gravitational backgrounds}

We come now to the discussion of GW backgrounds which are thought to be
produced at galactic and cosmic scales, though they have not yet been detected%
~\cite{Schutz,Maggiore}.
For simplicity, we restrict our attention to the case of a stationary, 
unpolarized and isotropic background, assuming uncorrelated noises
with the same noise energy in all the modes corresponding to a given
frequency. Furthermore, we study the classical limit where the number 
$n_\gw $ of gravitons per mode is much larger than unity (more detailed
discussion below).

The metric fluctuations are thus given by the graviton propagator~\cite{Jaekel94}
here written in the TT gauge and at the classical limit~\cite{Reynaud01}
\begin{eqnarray}
\label{metric_fluctuations}
&&<h_{ij}^\TT(x^\prime) h_{kl}^\TT(x)> = {32 \pi^2 G\hbar\over c^3} 
\int {d^4 k\over(2\pi)^4}\proj_{ijkl} n_\gw [|\omega|] \delta(k^2) e^{-ik(x^\prime - x)}
\end{eqnarray}
The spectral energy density of GW appearing in (\ref{metric_fluctuations}) 
has been defined as in \cite{Landau}. 
The fluctuations (\ref{metric_fluctuations}) lie on the light cone and 
$\proj_{ijkl}$ accounts for the sum over graviton helicity states
(see eq.~\ref{projtrans}).
The number $n_\gw$ of gravitons is a measure of the noise energy $e_\gw$
per mode at positive frequency $\omega$ or of an equivalent noise
temperature $T_\gw$ ($\kB$ is the Boltzmann constant)
\begin{equation}
e_\gw=\hbar \omega n_\gw=\kB T_\gw
\end{equation}
Let us stress immediately~\cite{Reynaud01} that $T_\gw$ is certainly not 
a thermodynamical temperature. 
It is rather an effective noise temperature of the GW environment, 
which is in fact extremely weakly coupled to other fields.
$T_\gw$ will turn out to have an enormous value, much larger than any equilibrium
temperature, and also to depend on frequency.
Only in special cases will the background correspond to a constant $T_\gw$, 
at least in some frequency range of interest (see below).
The GW noise corresponds to a classical limit $n_\gw\gg1$ and 
probably dominates quantum sources of gravity fluctuations. 
Anyway, if such sources are found which are 
thought to play a role for some phenomena, they have to be 
compared with the galactic and cosmic backgrounds discussed below,
which are direct consequences of current standard physics.

For the sake of comparison with the papers on GW detection \cite{Abbott}, 
we introduce a further notation, corresponding to the ratio $\Omega_\gw$ 
of GW spectral energy density to the cosmic closure density.
The relation with the characterization used in the present paper
is as follows (using the angular integrals written in \cite{Lamine02}) 
\beq
\label{cosmic_background}
&&C_{h_{ij}h_{kl}} [ \omega ]  = S_\gw[\omega] 
\left( {\delta_{ik}\delta_{jl}+\delta_{il}\delta_{jk} \over2}
-{\delta_{ij}\delta_{kl} \over3} \right)
\\
&&S_\gw[\omega] = \frac{32G\kB T_\gw}{5c^5} 
=\frac{3 H_0^2}{10\pi^2 f^3}\,\Omega_\gw
\nonumber
\eeq
$C_{h_{ij}h_{kl}} [ \omega ]$ represents the correlation 
of different metric components evaluated as functions of time 
at a fixed space position.
For example, the quantity involved in the calculation of the preceding 
section is 
\beq
&&C_{h_{11}h_{11}} [ \omega ]  = \frac23 S_\gw[\omega] 
= \frac{H_0^2}{5\pi^2 f^3}\,\Omega_\gw
\eeq
$f$ is the GW frequency and $H_{0}$ the Hubble constant measured
with the dimension of a frequency
(with the currently preferred value
$H_{0} \simeq 71\mathrm{\ km\ s}^{-1}\mathrm{\ Mpc}^{-1}$, we get
$H_{0} \simeq 2.3\times 10^{-18}\s^{-1}$). 

Present knowledge on GW backgrounds comes from studies estimating the 
probability of events which might be observed by interferometric GW
detectors. A first important contribution is constituted by the ``binary 
confusion background'', that is the background of GW emitted by unresolved 
binary systems in the galaxy and its vicinity~\cite{BCB}.
This `binary confusion background' leads to a nearly flat noise spectrum,
that is also to a nearly flat temperature, in the $\mu\Hz$ to mHz frequency range 
\beq
\label{gw_temperature}
&&10^{-6} \Hz \lesssim  \frac{\omega} {2\pi} \lesssim 10^{-3}\Hz \quad , \quad
S_\gw \sim 10^{-34} \Hz^{-1} \quad , \quad
T _\gw \sim 10^{41}\K
\eeq
This enormous value, much larger than Planck temperature ($\sim 10^{32}$K), 
certainly means that it does not correspond to an equilibrium temperature.
Previous estimations correspond to the confusion background 
of GW emitted by binary systems in our Galaxy or its vicinity. 
As a consequence of the large number of unresolved and independent sources, 
and of the central limit theorem, this stochastic noise   
should obey gaussian statistics. 

Besides the galactic background, there also exist predictions for GW backgrounds 
associated with a variety of cosmic processes \cite{Schutz,Maggiore},
which have a more speculative character but could constitute a new window
on primordial cosmology. 
The predictions for the parameter $\Omega_\gw $ are strongly model dependent.
They tend to produce nearly constant values for $\Omega_\gw $ 
which means that the noise spectrum $S_\gw$ or noise temperature $T_\gw$
increases rapidly when the frequency decreases, with large amounts
of noise at low frequencies down to Hubble frequency.
Cosmic noises do not correspond to stationary fluctuations since they are
usually produced by parametric amplification processes of primordial
fluctuations~\cite{Grishchuk}. 
Note also that the galactic background is probably not isotropic 
and might even not be unpolarized.
Though this means that our description of GW backgrounds misses some features
of our real GW environment,
we will however go on with this simple description.
Our aim will essentially be to discuss generic phenomena
such as diffusion and decoherence, and to shed some light
on their dependence on the mass or size of the studied systems.
It will become apparent later on that a qualitative description of the
environment is sufficient to this aim.

Before embarking on this discussion, let us refer the reader to
review papers which collect the constraints obtained on the GW noise levels
from a variety of observations.
Schutz~\cite{Schutz} and Maggiore~\cite{Maggiore} give a lot of
information, oriented in particular to the discussion of the 
space project LISA~\cite{GWVarenna}. 
Abbott \textit{et al}~\cite{Abbott} collect the bounds on $\Omega_\gw$
deduced in various frequency windows from a variety of observations
(see in particular their figure 14 and related explanations).

\section{Gravitational diffusion and decoherence in interferometers}

As already stated, it has recently been suggested that matter-wave interferometers 
could have enough sensitivity to gravitation fields to be able to reveal 
decoherence induced by spacetime or gravitation fluctuations%
~\cite{Percival,Garay98,Amelino,Bingham}.
The argument was one of the motivations for studying the atomic interferometer 
HYPER which was designed for measuring the Lense-Thirring effect in space%
~\cite{Hyper00,HYPER} and it remains present in new projects for even more 
sensitive atomic probes~\cite{CVproposals}. 
In the present section, we discuss the effect of GW on interferometers
and take the figures corresponding to HYPER in order to discuss 
the corresponding orders of magnitude~\cite{Lamine02,Reynaud04,Lamine06}.

The use of atomic interferometers as sensors of inertial and gravitational 
effects through a measurement of the dephasing between its two arms has 
been discussed in a number of papers (see for instance~\cite{Borde,Peters} 
for reviews) and in several lectures at this School~\cite{AtomsVarenna}. 
Here we choose to use a simple approach to the dephasing~\cite{Linet76},
analogous to the range (\ref{finiterange}),
and focus the discussion on the comparison between matter-wave and 
electromagnetic-wave interferometers.
To this aim, we first consider HYPER used as a gyrometer measuring
the rotation of the interferometer with respect to inertial
frames through the observation of a Sagnac effect.
The Sagnac dephasing $\Phi$ is proportional 
to the mass $m_\at$ of the (non relativistic) atoms, 
to the area $A$ of the interferometer
and to the rotation frequency $\Omega$ 
\beq
&&\Phi_\Sagnac = \frac{ 2 m_\at A} {\hbar} \Omega\quad,
\quad  A  = v_\at^2 \tau_\at^2 \sin\alpha
\eeq
$v_\at$ is the atomic velocity and $\tau_\at$ is the time of flight of atoms
on one HYPER side, so that the area $A$ is given by the length $v_\at \tau_\at$ 
of one side and the aperture angle $\alpha$.

The rotation of the instrument is measured with respect to the local inertial frame 
which differs from the celestial frame determined by pointing at ``fixed stars'',
as a consequence of the Lense-Thirring effect.
Such an effect, a dragging of inertial frames, or gravitomagnetic effect, 
induced by the neighbouring Earth, would be measured by HYPER, through the comparison of 
the reading of the atomic interferometer and of
the indications of a star tracker~\cite{Hyper00,HYPER}. 
Though the Lense-Thirring effect is dominated by the near field of the neighbour Earth,
it also feels contributions of other gravitating bodies.
The effect of GW discussed now can be regarded
as the far field effect of distant binaries. 

For calculating the effect of GW, we use an expression of the dephasing which is valid for 
matter-wave as well as electromagnetic-wave interferometers~\cite{Linet76}
(and which reduces in the latter case to the already written range variation \ref{finiterange})
\beq
&&\delta \Phi_\gw = {m_\at\over 2\hbar}\oint h_{\rm i j}^\TT v_\at^{\rm i} v_\at^{\rm j} \d\tau
= \frac{2m_\at A} {\hbar} \delta \Omega 
\label{Phigw}
\eeq
(the symbol $\oint$ denotes the difference between integrals over the two arms of the interferometer).
For a rhombic shape of the interferometer (supposed to lie in the spatial plane 12), 
this expression may be rewritten from the derivative of the metric component $h_{12}$ 
\beq
&&\delta\Omega (t) = -\frac 12 \frac{\d \overline{h_{12}^\TT}}{\d t}, \qquad
\overline{h_{12}^\TT}(t) = \int \ h_{12}^\TT \left(t - \tau \right) 
g\left(\tau\right) \d \tau
\eeq
The linear filtering function $g$ has
a triangular shape which reflects the distribution of the time of exposition 
of atoms to GW inside the rhombic interferometer.
The square of its Fourier transform, which describes linear filtering in frequency space, 
is an apparatus function  characterizing the interferometer
\beq
&&|\tilde{g} \left[ \omega \right]|^2 = 
\left( \frac { \sin \frac{\omega \tau_\at}{2} } 
{ \frac{\omega \tau_\at}{2} } \right) ^4 
\eeq

We then evaluate the fading of fringe contrast obtained by averaging 
over stochastic dephasings. 
This evaluation~\cite{Lamine02} can be shown to be equivalent
to the other approaches to decoherence (see for example~\cite{Imry90}). 
Stochastic GW with frequencies higher than the inverse of the averaging time
identify with the unobserved degrees of freedom which
are usually traced over in decoherence theory.
When $\delta \Phi_\gw$
is a gaussian stochastic variable, the degraded fringe contrast is read as 
\beq
&&\left\langle \exp \left( i\delta \Phi_\gw \right) \right\rangle =
\exp \left( -\frac{ \Delta \Phi_\gw^2 } {2} \right), \qquad
\Delta \Phi_{\gw}^2 = \left\langle \delta \Phi_\gw^2 \right\rangle 
\eeq
The variance $\Delta \Phi_\gw^2$ can be written as an integral over 
the noise spectrum 
\beq
&&C_{h_\mathrm{12}h_\mathrm{12}} [ \omega ]  = \frac12 S_\gw[\omega] 
\eeq
In the case of an approximately flat spectrum 
which, as already discussed, is approximately realized by the 
binary confusion background between the $\mu$Hz and mHz ranges, 
the variance is found to be proportional to the time of exposition $\tau_\at$
\beq
\label{micro_factor}
\Delta \Phi_{\gw}^2 = \left(\frac{2m_\at v_\at^2}{\hbar} \sin\alpha\right)^2\ S_\gw \tau_\at
\eeq
This means that the effect of stochastic GW is equivalent to a brownian diffusion
process induced by the white noise.

The other relevant parameters are the kinetic energy $\propto m_\at v_\at^2$ of the 
atomic probe and the aperture $\alpha$ of the interferometer.
After substitution  of the numbers corresponding to HYPER~\cite{Hyper00,HYPER}, 
we deduce that the decoherence is completely negligible
($\Delta \Phi^2_\gw \ll 1$).
Note that the effect of GW on the lasers involved in the stimulated Raman processes 
used for building up beam splitters and mirrors has also to be considered~\cite{Lamine02},
without changing the qualitative conclusion.
In fact, the phase diffusion induced by the scattering of GW remains
much smaller than the phase diffusion induced by  instrumental fluctuations,
for example mechanical vibrations of the mirrors.

\section{Classical versus quantum behaviours}	

We come to another case which lies at the opposite end of extremely macroscopic systems, 
namely Moon orbiting around Earth.
In this case, GW lead to an extremely rapid decoherence
process~\cite{Reynaud01}.
Moreover, although decoherence is usually attributed to collisions
of residual gaz, radiation pressure of solar radiation or, even, 
scattering of electromagnetic fluctuations in the cosmic microwave background,
we show that the decoherence of planetary motions is dominated by the 
scattering of stochastic GW.

The Earth and Moon constitute a binary system with a large quadrupole momentum,
so that its internal motion is highly sensitive to GW.
For the sake of simplicity, we describe the Earth-Moon system as a circular planetary 
orbit in the plane $x_1 x_2$. We also use the reduced mass $m$ (defined from the masses 
of the two bodies), the radius $\rho$ (the constant distance between the two masses), 
with the orbital frequency $\Omega$, the normal acceleration $a$ on the circular orbit 
and the tangential velocity $v$ related through 
\beq
\label{Kepler}
a = \rho \Omega ^2 = \frac {v ^2} \rho 
\eeq
The GW perturbation on the relative position $x^{\rm i}$ 
in the binary system amounts to a tidal force $\delta F$
which may also be seen as a geodesic deviation 
\beq
\label{gw_force}
&&\delta \dot{p}_{\rm i}(t) = \delta F_{\rm i}(t) = m c^2 R_{\rm 0 i 0 j} x^{\rm j}(t)
\eeq

The stochastic GW background then induces a Brownian motion on the relative 
position of the Moon, which  may be characterized by a momentum diffusion 
with a variance varying linearly with the time of exposition $\tau$ 
\beq
\label{Brownian}
&&<\delta p^2(t)> = 2 D _\gw \tau 
\eeq
The momentum diffusion coefficient $D_\gw$ is obtained~\cite{Reynaud01}
with the form of a fluctuation-dissipation relation~\cite{brownian} 
\beq
\label{gw_damping}
&&D_\gw = m \Gamma _\gw \kB T_\gw\quad, \quad \Gamma _\gw =\frac{32Gma^2}{5c^5}
\eeq
$T_\gw$ is the effective noise temperature of the GW background,
evaluated at twice the orbital frequency, and 
$\Gamma _\gw$ is a damping rate which
corresponds to another well-known Einstein formula, namely
the quadrupole formula for GW emission \cite{Einstein18}.
Gravitational damping is extremely small for the Earth-Moon system,
in fact much smaller than the damping due to other 
environmental fluctuations, such as electromagnetic radiation pressure or
Earth-Moon tides, the latter being the dominant contribution
to damping~\cite{Bois96}
\beq
\label{Moon_damping}
&&\Gamma _{\gw} \approx 10^{-34}\ \s^{-1} \ll \ \Gamma _{\rm em} \ < \ \Gamma _{\rm tides}
\eeq

In contrast, the GW contribution to decoherence appears to be much larger 
than the contributions associated with tide interactions and electromagnetic
scattering
\beq
&&D_\gw \ \gg \ D_{\rm tides} \ > \ D_{\rm em} 
\eeq
This change of hierarchy results from the dependence of diffusion
on the noise level, that is on the noise temperature.
While the ratio $\frac{\Gamma_\gw}{\Gamma_{\rm tides}}$
of the damping constants associated with gravitational waves and
tides is of the order of $10^{-16}$, the ratio $\frac{T_\gw}{T_{\rm tides}}$
of the temperatures is of the order of $10^{38}$.
It follows that the ratio $\frac{D_\gw}{D_{\rm tides}}$
of the diffusion constants is very large so that the GW contribution to
decoherence dominates the other ones.

Decoherence can be evaluated by considering two neighbouring 
internal motions of the planetary system which correspond to the same
spatial geometry but slightly different values of the epoch, the time of passage at a given space point.
For simplicity, we measure this difference by the spatial distance $\Delta x$ 
between the two motions, which is constant for uniform motion.
The variation of momentum (\ref{gw_force}) results in a perturbation of the quantum phase 
one may associate with the relative position in the binary system
\beq
\delta\Omega(t)  = {\delta p_{\rm i}(t)\over\hbar}\Delta x^{\rm i}
\eeq
The difference of phase between two neighboring motions then undergoes a Brownian motion \cite{Reynaud01},
resulting in a random exponential factor $e ^{i \delta \Phi_\gw}$.
Averaging this quantity over the stochastic effect of 
gravitational waves, still supposed to obey gaussian statistics, 
one obtains a decoherence factor 
\beq
&&\left\langle e ^{i \delta \Phi_\gw} \right\rangle 
= \exp \left( -\frac {\Delta \Phi_\gw^2} 2 \right) 
\eeq
The decoherence factor may be expressed in terms of the variables characterizing
the Brownian motion (\ref{Brownian}) and the distance between two motions $\Delta x$
\beq
\label{gw_decoherence}
&&\Delta \Phi_{\gw}^2 = \frac{2 D _\gw \Delta x^2 \tau} {\hbar ^2} 
\eeq
Relation (\ref{gw_decoherence}) agrees with the result expected from 
general discussions on decoherence~\cite{Zurek81}: decoherence 
efficiency increases exponentially fast with $\tau$ and $\Delta x^2$.

Using the numbers corresponding to the Earth-Moon system, one finds that 
decoherence is extremely efficient 
\beq
&&\frac{D_\gw}{\hbar^2} \approx 10^{75}\ \s^{-1}\m^{-2} 
\eeq
For $\Delta x$ as small as the Planck length, coherences are lost
on a very short time $\sim10\mu$s.
It follows that a macroscopic system such as the Earth-Moon system
can, for all purposes, be considered as classical.
We want to emphasize at this point that the dominant mechanism leading to 
this classical behavior is the scattering of GW
constituting our gravitational environment.
It is remarkable that the classicality and the ultimate fluctuations 
of very macroscopic systems appear to be determined by the classical 
gravitation theory which also explains their mean motion. 

\section*{Conclusion}

The results obtained in the previous sections for GW-induced decoherence are 
reminiscent of the qualitative discussions of the Introduction.
For microscopic probes, such as the atoms or photons involved in interferometers,
decoherence is so inefficient that it can be ignored so
that ordinary quantum mechanics is the appropriate description.
For large macroscopic bodies in contrast, decoherence is so efficient 
that quantum coherences between different positions are never observed, 
leading to the possibility of a purely classical description.

In both the microscopic and macroscopic cases, 
the decoherence factors $e^{-{\Delta \Phi_\gw ^2 \over2}}$ take similar forms. 
They may indeed be written in terms of a phase diffusion variance
depending on a few relevant factors \cite{Lamine02,Reynaud04,Lamine06}
\beq
&&\frac{\Delta \Phi_\gw ^2}{2} \simeq \left(\frac{mv^2\sin\alpha}{\mP c^2} \right)^2 \ 
\Theta _\gw \tau \quad, \quad \Theta _\gw  \simeq \frac{\kB T_\gw }{\hbar} 
\label{DeltaPhi2}
\eeq
The ratio $\frac{m^2}{\mP^2}$ confirms the preliminary arguments of the Introduction, 
namely that the Planck mass effectively plays a role in the definition of
a borderline between microscopic and macroscopic masses. 
However, other factors in the formula imply that the scaling argument on masses 
is not sufficient to obtain quantitative estimates. 
The presence of the probe velocity over light velocity implies that 
the parameter to be compared with Planck energy $\mP c^2$ is the
kinetic energy $m v^2$ of the probe rather than its mass energy $m c^2$.
Geometry also plays a role with the equivalent aperture angle $\alpha$ 
appearing in the formula.
Finally, the GW noise level is characterized by $\Theta _\gw $,
the temperature of the background measured as a frequency
(with $\Theta _\gw \sim 10^{52}\s^{-1}$ on the plateau of the galactic background). 
This very large value suggests that the transition between quantum and classical
behaviors could in principle be observed for masses smaller than Planck mass.

Formula (\ref{DeltaPhi2}) allows one to discuss whether or not the quantum/classical transition 
induced by intrinsic gravitational fluctuations could be observed~\cite{Lamine06}. 
It clearly favors experiments using heavy and fast particles in interferometers. 
Though interference patterns have been observed on rather large molecules%
~\cite{Zeilinger}, one checks that the numbers in these experiments are such 
that the GW-induced quantum decoherence remains negligible. 
An attractive alternative is to consider interferometers using quantum condensates%
~\cite{condensates}, an approach requiring further technological progress.
Note that phase diffusion could in principle be seen in interferometers
long before decoherence takes place, which lets free room for ideas 
which could allow one to observe the GW fluctuations acting on quantum systems.

\acknowledgments
Thanks are due for stimulating discussions to A. Lambrecht, P.A. Maia Neto,
R. Hervé, G. Ingold, L. Duchayne, P. Wolf, 
the members of the \textsc{Hyper}~\cite{HYPER}, \textsc{Sagas}~\cite{SAGAS} 
\textsc{Mwxg} and \textsc{Gauge} teams~\cite{CVproposals}.

\def\etal{\textit{et al }}
\def\ibid{\textit{ibidem }}
\def\url#1{\textrm{#1}}
\def\arxiv#1{\textrm{#1}}
\def\REVIEW#1#2#3#4{\textit{#1} \textbf{#2} {#4} ({#3})}
\def\Book#1#2#3{\textit{#1} ({#2}, {#3})}
\def\BOOK#1#2#3#4{\textit{#1} ({#2}, {#3}, {#4})}
\def\BOOKed#1#2#3#4#5{\textit{#1}, #2 ({#3}, {#4}, {#5})}
\def\Name#1#2{\textsc{#1}~#2}

\end{document}